\documentclass[pre, showkeys, nofootinbib, floatfix]{revtex4}
\usepackage{latexsym}
\usepackage{amsmath}
\usepackage[dvips]{graphicx}

\begin {document}
\title{Late comment on Astumian's paradox}
\author{L. \surname{Pal}}


\affiliation{KFKI Atomic Energy Research Institute H-1525 Budapest
114, POB 49 Hungary}

\begin{abstract}
In 2001 Astumian \cite{ast01} published a very simple game which
can be described by a Markov chain with absorbing initial and
final states. In August 2004 Piotrowski and Sladowski \cite{pio04}
asserted that Astumian's analysis was flawed. However, as was
shown by Astumian \cite{ast04}, this statement was wrong. In this
comment the properties of Markov chains corresponding to games
that are more general than that studied by Astumian, are
investigated.
\end{abstract}


\maketitle

\section{Introduction}

The present note was initiated by the revisited Astumian's
paradox. In August 2004 Piotrowski and Sladowski \cite{pio04}
asserted that Astumian's analysis was flawed. However, as shown by
Astumian \cite{ast04}, this statement was wrong. Since the
analysis of the problem in a slightly more general frame than it
was done earlier could be a good exercise for graduate students,
we came to the conclusion that it might be useful to publish our
elementary considerations about the properties of Markov chains
corresponding to  Astumian type games.

For entirely didactic reasons, in Sections II and III we present a
brief summary of definitions and statements which are needed for
the analysis of the Astumian type Markov chains. In Section IV we
analyze the properties of such chains and determine the
probabilities of losing and winning. Conclusions are made in
Section V.

\section{Preliminaries}

Let ${\mathcal N} = \{1,2,\ldots,N\}$ be a finite set of positive
integers, and ${\mathcal Z} = \{0,1,\ldots\}$ be a set of
non-negative integers. Denote by $\xi_{n},\; n \in {\mathcal Z}$
the random variable which assumes the elements of ${\mathcal N}$.
We say that the sequence $\{\xi_{n}\}$ forms a Markov chain if for
all $n \in {\mathcal Z}$ and for all possible values of random
variables the equation
\begin{equation} \label{1}
{\mathcal P}\{\xi_{n}=j \vert \xi_{0}=i_{0}, \xi_{1}=i_{1},
\ldots, \xi_{n-1}=i_{n-1}\} = {\mathcal P}\{\xi_{n}=j \vert
\xi_{n-1}=i_{n-1}\}
\end{equation}
is fulfilled. If $\xi_{n}=j$ then the process is said to be in
\textit{state} ${\mathcal S}_{j}$ at the $n$th (discrete time
instant) step. The states ${\mathcal S}_{1}, {\mathcal S}_{2},
\ldots, {\mathcal S}_{N}$ define \textit{the space of states} of
the process. The probability distribution ${\mathcal
P}\{\xi_{0}=i\}, \; i \in {\mathcal N}$ of the random variable
$\xi_{0}$ is called the \textit{initial distribution} and the
conditional probabilities ${\mathcal P}\{\xi_{n}=j \vert
\xi_{n-1}=i\}$ are called \textit{transition probabilities}. If
$\xi_{n-1}=i$ and $\xi_{n}=j$, then we say that the process made a
transition ${\mathcal S}_{i} \rightarrow {\mathcal S}_{j}$ at the
$n$th step. The Markov chain is \textit{homogeneous} if the
transition probabilities are independent of $n \in {\mathcal Z}$.
In this case we may write
\[ {\mathcal P}\{\xi_{n}=j \vert \xi_{n-1}=i\} = w_{ij}(1) = w_{ij}, \]
and it obviously holds that
\begin{equation} \label{2}
\sum_{j=1}^{N} w_{ij} = 1, \;\;\;\;\;\; \forall \; i \in {\mathcal
N}.
\end{equation}
In what follows we shall consider only homogeneous Markov chains.
We would like to emphasize that the transition probability matrix
\begin{equation} \label{3}
\mathbf{w} =
\begin{pmatrix}
w_{11} & w_{12} & \cdots & w_{1N} \\ w_{21} & w_{22} & \cdots &
w_{2N} \\ \vdots & \vdots & \ddots & \vdots \\ w_{N1} & w_{N2} &
\cdots & w_{NN}
\end{pmatrix},
\end{equation}
which is a \textit{stochastic matrix}, and the initial
distribution $p_{i} = {\mathcal P}\{\xi_{0}=i\}, \; i \in
{\mathcal N}$ determine the random process uniquely. For the sake
of simplicity, we assume that the process is a random walk of an
abstract object, called \textit{particle} on the space of states
${\mathcal S}_{1}, {\mathcal S}_{2}, \ldots, {\mathcal S}_{N}$.
The $n$th step transition probability
\begin{equation} \label{4}
{\mathcal P}\{\xi_{m+n}=j \vert \xi_{m}=i\} = w_{ij}(n)
\end{equation}
satisfies the following equation:
\begin{equation} \label{5}
w_{ij}(n) = \sum_{k=1}^{N} w_{ik}(r)\;w_{kj}(s),
\end{equation}
where
\[ r + s = n. \]
It is to note that $w_{ij}(n)$ is the probability that at the
$n$th step the particle is in the state ${\mathcal S}_{j}$
provided that at $n=0$ it was in the state ${\mathcal S}_{i}$.
From Eq. (\ref{5}) we obtain that
\[ w_{ij}(n) = \sum_{k=1}^{N} w_{ik}\;w_{kj}(n-1) = \sum_{k=1}^{N}
w_{ik}(n-1)\;w_{kj}, \] and by using the rules of matrix
multiplication we arrive at
\begin{equation} \label{6}
{\mathbf w}(n) = {\mathbf w} \; {\mathbf w}(n-1) = {\mathbf
w}(n-1)\;{\mathbf w} = {\mathbf w}^{n},
\end{equation}
where
\begin{equation} \label{7}
\mathbf{w}(n) =
\begin{pmatrix}
w_{11}(n) & w_{12}(n) & \cdots & w_{1N}(n) \\ w_{21}(n) &
w_{22}(n) & \cdots & w_{2N}(n) \\ \vdots & \vdots & \ddots &
\vdots
\\ w_{N1}(n) & w_{N2}(n) & \cdots & w_{NN}(n)
\end{pmatrix},
\end{equation}
and
\[ \mathbf{w}(0) =
\begin{pmatrix}
1 & 0 & \cdots & 0 \\ 0 & 1 & \cdots & 0
\\ \vdots & \vdots & \ddots & \vdots
\\0 & 0 & \cdots & 1
\end{pmatrix}\]
is the $N \times N$ unit matrix.

Making use of the total probability theorem we can determine
\textit{the absolute probabilities} $p_{j}(n)$ as follows:
\begin{equation} \label{8}
p_{j}(n) = \sum_{i=1}^{N} p_{i}\;w_{ij}(n), \;\;\;\;\;\; j \in
{\mathcal N},
\end{equation}
where $p_{i} = {\mathcal P}\{\xi_{0}=i\}$ is the initial
probability. Clearly, $p_{j}(n)$ is the probability that the
particle is in the state ${\mathcal S}_{j}$ at the $n$th step.
Introducing the row vector
\begin{equation} \label{9}
\vec{p} = \{p_{1}, p_{2}, \ldots, p_{N}\}
\end{equation}
Eq. (\ref{8}) can be rewritten in the form:
\begin{equation} \label{10}
\vec{p}(n) = \vec{p}\;{\mathbf w}(n) = {\mathbf
w}^{(T)}(n)\;\vec{p}^{(T)},
\end{equation}
where the upper index $T$ indicates the transpose of matrix
${\mathbf w}(n)$ and vector $\vec{p}$ defined by (\ref{7}) and
(\ref{9}), respectively. If the process starts from the state
${\mathcal S}_{i}$, then
\[ \vec{p} = \{0_{1}, 0_{2}, \ldots, 1_{i}, \ldots, 0_{N}\}, \]
and
\[ \vec{p}_{i}(n) = \{w_{i1}(n), w_{i2}(n),\ldots, w_{iN}(n)\}.\]

\section{Types of states and asymptotic behavior}

\subsection{Basic definitions}

In order to use clear notions, we introduce several well-known
definitions. If there is an integer $n \geq 0$ such that
$w_{jk}(n) > 0$, then we say the state ${\mathcal S}_{k}$ can be
reached from the state ${\mathcal S}_{j}$. If ${\mathcal S}_{k}$
can be reached from ${\mathcal S}_{j}$ and ${\mathcal S}_{j}$ can
be reached from ${\mathcal S}_{k}$, then ${\mathcal S}_{j}$ and
${\mathcal S}_{k}$ are \textit{connected states}. Obviously, if
${\mathcal S}_{j}$ and ${\mathcal S}_{k}$ are not connected, then
either $w_{jk}(n) = 0$, or $w_{kj}(n) = 0$. The set of states
which are connected forms a \textit{class of equivalence}. A
Markov chain is called \textit{irreducible} if every state can be
reached from every state i.e., the entire state space consists of
only one class of equivalence. In other words, the Markov chain is
irreducible when all of the states are connected.

The probability $f_{ij}(n)$ of passage from ${\mathcal S}_{i}$ to
${\mathcal S}_{j}$ in exactly $n$ steps, that is, without passing
through ${\mathcal S}_{j}$ before the $n$th step, is given by
\begin{equation} \label{11}
f_{ij}(n) = \sum_{j_{1} \neq j,\; j_{2} \neq j, \;\ldots \;,
j_{n-1} \neq j} w_{ij_{1}}\;w_{j_{1}j_{2}} \; \cdots
\;w_{j_{n-1}j}.
\end{equation}
There exists an important relationship between the probabilities
$w_{ij}(n)$ and $f_{ij}(n)$ which is easy to prove.  The
relationship is given by
\begin{equation} \label{12}
w_{ij}(n) = \sum_{k=1}^{n} f_{ij}(k)\;w_{jj}(n-k), \;\;\;\;\;\;
\forall \; n \in {\mathcal Z}.
\end{equation}
One has to note that the expressions $w_{jj}(0) = 1$ are the
diagonal elements of the unit matrix ${\mathbf w}(0)$.

The proof of (\ref{12}) is immediate upon applying the total
probability rule. The particle passes from ${\mathcal S}_{i}$ to
${\mathcal S}_{j}$ in $n$ steps if, and only if, it passes from
${\mathcal S}_{i}$ to ${\mathcal S}_{j}$ for the first time in
exactly $k$ steps, $k = 1, 2, \ldots, n$, and then passes from
${\mathcal S}_{j}$ to ${\mathcal S}_{j}$ in the remaining $n-k$
steps. These ``paths" are disjoint events, and their probabilities
are given by $f_{ij}(k)\;w_{jj}(n-k)$. Summing over $k$ one
obtains the equation (\ref{12}).

Let us introduce the generating functions
\begin{equation} \label{13}
\varphi_{ij}(z) = \sum_{n=1}^{\infty} f_{ij}(n)\;z^{n}
\;\;\;\;\;\; \text{and} \;\;\;\;\;\; \omega_{ij}(z) =
\sum_{n=1}^{\infty} w_{ij}(n)\;z^{n}.
\end{equation}
Taking into account that $w_{jj}(0) = 1$,  from Eq. (\ref{12}) we
obtain
\begin{equation} \label{14}
\omega_{ij}(z) = \varphi_{ij}(z)\left[1 + \omega_{jj}(z)\right],
\end{equation}
and from this
\begin{equation} \label{15}
\varphi_{ij}(z) = \frac{\omega_{ij}(z)}{1 + \omega_{jj}(z)},
\end{equation}
so we have
\begin{equation} \label{16}
F_{ij} = \sum_{n=1}^{\infty} f_{ij}(n) = \frac{\sum_{n=1}^{\infty}
w_{ij}(n)}{1 + \sum_{n=1}^{\infty} w_{jj}(n)},
\end{equation}
and in particular
\begin{equation} \label{17}
\sum_{n=1}^{\infty} w_{jj}(n) = \frac{\sum_{n=1}f_{jj}(n)}{1 -
\sum_{n=1}^{\infty}f_{jj}(n)}.
\end{equation}

$F_{ij}$ defined by (\ref{16}) is the probability that a particle
starting its walk from ${\mathcal S}_{i}$ passes through the state
${\mathcal S}_{j}$ at least once. Clearly, $F_{ii} = F_{i}$ is
\textit{the probability of returning to ${\mathcal S}_{i}$ at
least once}.

More generally, the probability $F_{ij}(k)$ that a particle
starting its walk from ${\mathcal S}_{i}$ passes through
${\mathcal S}_{j}$ \textit{at least $k$ times} is given by
\[ F_{ij}(k) = \left[\sum_{n=1}^{\infty}
f_{ij}(n)\right]\;F_{jj}(k-1) = F_{ij}\;F_{jj}(k-1). \] In
particular, the probability of returning to ${\mathcal S}_{i}$ at
least $k$ times is given by $F_{ii}(k) = (F_{ii})^{k}$. Its limit
\[ R_{ii} = \lim_{k \rightarrow \infty} (F_{ii})^{k} = \left\{
\begin{array} {ll}
0, & \text{if $F_{ii} < 1$,} \\
\text{ } & \text{ } \\
1, & \text{if $F_{ii} = 1$}
\end{array} \right. \]
is the probability of \textit{returning to ${\mathcal S}_{i}$
infinitely often}. It follows from the previous relationship that
the probability that a particle starting its walk from ${\mathcal
S}_{i}$ passes through ${\mathcal S}_{j}$ infinitely many times is
\[ R_{ij} = \lim_{k \rightarrow \infty} F_{ij}(k) =
F_{ij}\;R_{jj}, \] so that
\[ R_{ij} = \left\{
\begin{array} {ll}
0, & \text{if $F_{ii} < 1$,} \\
\text{ } & \text{ } \\
F_{ij}, & \text{if $F_{ii} = 1$.}
\end{array} \right. \]

We say that ${\mathcal S}_{i}$ is a \textit{return state} or a
\textit{nonreturn state} according as $F_{i} > 0$ or $F_{i} = 0$.
As a further definition, we say that ${\mathcal S}_{i}$ is a
\textit{recurrent state} or a \textit{nonrecurrent state}
according as $F_{i} = 1$ or $0 \leq F_{i} < 1$. A nonrecurrent
state is often called a \textit{transient state}.

The state ${\mathcal S}_{i}$ is called \textit{periodic} with
period $\ell$ if a return to ${\mathcal S}_{i}$ can occur only at
steps $\ell, 2 \ell, 3 \ell, \ldots $ and $\ell > 1$ is the
greatest integer with this property. If $n$ is not divisible by
$\ell$, then $w_{ij}(n) = 0$. If the period of each state is equal
to $1$, i.e., if $\ell = 1$, then the Markov chain is called
\textit{aperiodic}. In the sequel we are dealing with aperiodic
Markov chains.

A set ${\mathcal C}$ of states in a Markov chain is
\textit{closed} if it is impossible to move out from any state of
${\mathcal C}$ to any state outside ${\mathcal C}$ by one-step
transitions, i.e., $w_{ij}(1) = w_{ij} = 0$ if ${\mathcal S}_{i}
\in {\mathcal C}$ and ${\mathcal S}_{j} \not\in {\mathcal C}.$ In
this case $w_{ij}(n) = 0$ obviously holds for every $n \in
{\mathcal Z}$. If a single state ${\mathcal S}_{i}$ forms a closed
set, then we call this an \textit{absorbing state}, and we have
$w_{ii} = 1$.

The states of a closed set ${\mathcal C}$ are recurrent states
since the return probability $F_{i}$ for any state ${\mathcal
S}{_i} \in {\mathcal C}$ is equal to $1$. Therefore, the
\textit{set of recurrent states} is denoted by ${\mathcal C}$.
~\footnote{The set ${\mathcal C}$ can be decomposed into mutually
disjoint closed sets ${\mathcal C}_{1}, {\mathcal C}_{2}, \ldots,
{\mathcal C}_{r}$ such that from any state of a given set all
states of that set and no others can be reached . States
${\mathcal C}_{1}, {\mathcal C}_{2}, \ldots, {\mathcal C}_{r}$ can
be reached from ${\mathcal T}$, but not conversely.} The set of
states having return probabilities $F_{i} < 1$ is the \textit{set
of transient states} and it is denoted by ${\mathcal T}$.
Obviously, if ${\mathcal S}_{i} \in {\mathcal T}$ and ${\mathcal
S}_{j} \in {\mathcal C}$, i.e., if ${\mathcal S}_{j}$ is an
absorbing state, then $F_{ij}$ is the probability that a particle
starting at ${\mathcal S}_{i}$ is finally absorbed at ${\mathcal
S}_{j}$.

Let $\nu_{ij}$ be the \textit{passage time} of a particle from the
state ${\mathcal S}_{i}$ to the state ${\mathcal S}_{j}$, taking
values $m = 1, 2, \ldots, $ with probabilities $f_{ij}(m)$. If
\[ \sum_{m=1}^{\infty} f_{ij}(m) = F_{i,j} = 1, \]
then the \textit{expected passage time} $\tau_{ij} = {\mathbf
E}\{\nu_{ij}\}$ from ${\mathcal S}_{i}$ to ${\mathcal S}_{j}$ is
defined by
\[ \tau_{ij} = \sum_{m=1}^{\infty} m\;f_{ij}(m) =
\left[\frac{d\varphi_{ij}(z)}{dz}\right]_{z=1}, \] while if
$F_{ij} < 1$, one says that $\nu_{ij} = \infty$ with probability
$1 - F_{ij}$, i.e., if $F_{ij} < 1$, then the expected passage
time $\tau_{ij} = \infty$. If the state ${\mathcal S}_{j} =
{\mathcal S}_{i}$ and it is recurrent, i.e., if $F_{ii} = F_{i} =
1$, then the expectation
\begin{equation} \label{18}
{\mathbf E}\{\nu_{ii}\} = \sum_{m=1}^{\infty} m\;f_{ii}(m) =
\left[\frac{d\varphi_{ii}(z)}{dz}\right]_{z=1} = \tau_{ii} =
\mu_{i}
\end{equation}
is called \textit{mean recurrent time}. If $\mu_{i} = \infty$,
then we say that ${\mathcal S}_{i}$ is a \textit{recurrent
null-state}, whereas if $\mu_{i} < \infty$, then we say that
${\mathcal S}_{i}$ is a \textit{recurrent non-null-state}. If
$F_{i} < 1$, i.e., the state ${\mathcal S}_{i}$ is transient, then
$1 - F_{i}$ is the probability that the recurrence time is
infinitely long, and so $\mu_{i} = \infty$.

We say that the recurrent state ${\mathcal S}_{i}$ is
\textit{ergodic}, if it is not a null-state and is aperiodic, that
is, if $F_{i} = 1, \; \mu_{i} < \infty$ and $\ell = 1$.

\subsection{Asymptotic behavior}

The first statement is very simple, hence it is given without
proof. If ${\mathcal S}_{j}$ is a transient or a recurrent
null-state, then for any arbitrary ${\mathcal S}_{i}$
\begin{equation} \label{19}
\lim_{n \rightarrow \infty} w_{ij}(n) = 0
\end{equation}
holds.

If ${\mathcal S}_{i}$ and ${\mathcal S}_{j}$ are recurrent
aperiodic states due to the same closed set, then
\begin{equation} \label{20}
\lim_{n \rightarrow \infty} w_{ij}(n) = \frac{1}{\mu_{j}},
\end{equation}
irrespective of ${\mathcal S}_{i}$. ~\footnote{In order to prove
the limit relationship (\ref{20}) Tauber's Theorem is used instead
of the lemma by Erd\H os-Feller-Kac.}

If $i=j$, then we have from Eq. (\ref{14}) the formula
\begin{equation} \label{21}
\omega_{jj}(z) = \frac{\varphi_{jj}(z)}{1 - \varphi_{jj}(z)}.
\end{equation}
Substituting this into (\ref{14}) we obtain the following
expression:
\begin{equation} \label{22}
\omega_{ij}(z) = \varphi_{ij}(z)\left(1 + \frac{\varphi_{jj}(z)}{1
- \varphi_{jj}(z)}\right) = \frac{\varphi_{ij}(z)}{1 -
\varphi_{jj}(z)}.
\end{equation}
By using Tauber's Theorem we can state that
\begin{equation} \label{23}
\lim_{z \uparrow 1}\;(1-z)\;\frac{\varphi_{ij}(z)}{1 -
\varphi_{jj}(z)} = \lim_{n \rightarrow \infty} w_{ij}(n).
\end{equation}
Since ${\mathcal S}_{i}$ and ${\mathcal S}_{j}$ are aperiodic
recurrent states due to the same closed set,
\[ \lim_{z \uparrow 1}\varphi_{ij}(z) = \lim_{z \uparrow
1}\varphi_{jj}(z) = 1, \] i.e., the limit value we have to
determine
\[ \lim_{z \uparrow 1} \frac{1 - z}{1 - \varphi_{jj}(z)}. \]
Applying L'Hospital's rule we find that
\[ \lim_{z \uparrow 1} \frac{1 - z}{1 - \varphi_{jj}(z)} =
\frac{1}{\varphi'(1)} = \frac{1}{\mu_{j}}, \] and thus we obtain
(\ref{20}). This completes the proof.

As a generalization we would like to consider the case when
${\mathcal S}_{i}$ is a transient state (${\mathcal S}_{i} \in
{\mathcal T}$) and ${\mathcal S}_{j}$ is an aperiodic recurrent
state due to the closed set ${\mathcal C}$. It can be shown that
\begin{equation} \label{24}
\lim_{n \rightarrow \infty} w_{ij}(n) = \frac{F_{ij}}{\mu_{j}},
\end{equation}
where $F_{ij}$ is the probability that a particle starting from
${\mathcal S}_{i}$ will ultimately reach and stay in the state
${\mathcal S}_{j} \in {\mathcal C}$. In other words, $F_{ij}$ is
the \textit{absorption probability} that satisfies the following
system of equations:
\begin{equation} \label{25}
F_{ij} = w_{ij} + \sum_{{\mathcal S}_{k} \in {\mathcal T}}
w_{ik}F_{kj}, \;\;\;\;\;\; \forall \; {\mathcal S}_{i} \in
{\mathcal T}.
\end{equation}
Clearly, if ${\mathcal T} \cup {\mathcal C}$ contains all of the
possible states of the particle, then
\begin{equation} \label{26}
\sum_{{\mathcal S}_{j} \in {\mathcal C}} F_{ij} = 1.
\end{equation}
The proof of (\ref{24}) follows immediately from (\ref{22}). Since
\[ \lim_{z \uparrow 1}\varphi_{ij}(z)\;\frac{1 -z}{1 -
\varphi_{jj}(z)} = \sum_{n=1}^{\infty}
f_{ij}(n)\;\frac{1}{\mu_{j}} = \frac{F_{ij}}{\mu_{j}}, \] we
obtain the limit relationship (\ref{24}).

Finally, we would like to present a brief classification of Markov
chains.
\begin{itemize}
\item A Markov chain is called \textit{irreducible} if and only if all its
states form a closed set and there is no other closed set
contained in it.
\item A Markov chain is called \textit{ergodic} if the probability
distributions
\[ p_{j}(n) = \sum_{k=1}^{N} p_{k}(0)\;w_{kj}(n), \;\;\;\;\;\; j
\in {\mathcal N} \] always converge to a limiting distribution
$p_{j}$ which is independent of the initial distribution
$p_{j}(0)$, that is, when $\lim_{n \rightarrow \infty} p_{j}(n) =
p_{j}, \;\; \forall \; j \in {\mathcal N}$. All states of a
finite, aperiodic irreducible Markov chain are ergodic.
\item The probability distribution $p_{i}^{(st)}$ is a
\textit{stationary} distribution of a Markov chain if, when we
choose it as an initial distribution all the distributions
$p_{i}(n)$ will coincide with $p_{i}^{(st)}$. Every stationary
distribution of a Markov chain satisfies the following system of
linear equations:
\[ p_{j}^{(st)} = \sum_{i} p_{i}^{(st)}\;w_{ij} \;\;\;\;\;\;
\text{and} \;\;\;\;\;\; \sum_{j} p_{j}^{(st)} = 1, \] and
conversely, each solution $p_{j}^{(st)}$ of this system is a
stationary distribution of the Markov chain, if it is a
probability distribution.
\end{itemize}

It is to mention that some parts of this short summary is based on
the small but excellent book by Tak\'acs \cite{tak60}.

\section{Markov chains with absorbing states}

In this section we are going to deal with Markov chains containing
\textit{two absorbing states} ${\mathcal S}_{1}$ and ${\mathcal
S}_{N}$,  and $N-2$ \textit{transient states}. In this case, the
Markov chain is \textit{reducible} and \textit{aperiodic}. The set
of its states is the union of \textit{two closed sets} ${\mathcal
C}_{1} =  \{{\mathcal S}_{1}\}$ and ${\mathcal C}_{2} =
\{{\mathcal S}_{N}\}$, and of the set of transient states
${\mathcal T} = \{{\mathcal S}_{2}, {\mathcal S}_{3}, \ldots,
{\mathcal S}_{N-1}\}$  The states ${\mathcal S}_{1}$ and
${\mathcal S}_{N}$ can be reached from each state of ${\mathcal
T}$ but the converse doesn't hold, no state of ${\mathcal T}$ can
be reached from the states ${\mathcal S}_{1}$ and ${\mathcal
S}_{N}$. The states of ${\mathcal T}$ are \textit{non-recurrent}
since the particle leaves the set never to return to it. In
contrary, the states of ${\mathcal C}_{1}$ and ${\mathcal C}_{2}$
are ergodic.

\subsection{Chains of five states}

Let us assume that the transition matrix $\mathbf{w}$ has the
following form:
\begin{equation} \label{27}
\mathbf{w} =
\begin{pmatrix}
1 & 0 & 0 & 0 & 0 \\ w_{21} & w_{22} & w_{23} & 0 & 0 \\
0 & w_{31} & w_{33} & w_{34} & 0 \\
0 & 0 & w_{43} & w_{44} & w_{45} \\
0 & 0 & 0 & 0 & 1
\end{pmatrix},
\end{equation}
where
\[ \sum_{j} w_{ij} = 1. \]
The particle, which starts his walk from one of the states
${\mathcal S}_{i}; \;\; (i=2,3,4)$, is captured when it enters the
states ${\mathcal S}_{1}$ or ${\mathcal S}_{5}$. By using the
foregoing formulae for $F_{i1}$ and $F_{i5}$, we can immediately
obtain the capture probabilities by the absorbing states
${\mathcal S}_{1}$ and ${\mathcal S}_{5}$, respectively. In order
to have a direct insight into the nature of the process, we derive
the backward equations for the probabilities $w_{ij}(n)$. Clearly,
\begin{eqnarray}
w_{1j}(n) & = & \delta_{1j}, \label{28}\\
w_{2j}(n) & = & w_{21}\;w_{1j}(n-1) + w_{22}\;w_{2j}(n-1) +
w_{23}\;w_{3j}(n-1), \label{29} \\
w_{3j}(n) & = & w_{32}\;w_{2j}(n-1) + w_{33}\;w_{3j}(n-1) +
w_{34}\;w_{4j}(n-1), \label{30} \\
w_{4j}(n) & = &  w_{43}\;w_{3j}(n-1) + w_{44}\;w_{4j}(n-1) +
w_{45}\;w_{5j}(n-1), \label{31} \\
w_{5j}(n) & =& \delta_{5j}, \label{32}
\end{eqnarray}
and by introducing the generating function
\begin{equation} \label{33}
g_{ij}(z) = \delta_{ij} + \sum_{n=1}^{\infty} w_{ij}(n) \; z^{n} =
\delta_{ij} + \omega_{ij}(z), \;\;\;\;\;\;\; \vert z \vert < 1,
\end{equation}
we obtain the following system of equations:
\begin{eqnarray}
g_{1j}(z) & = & \delta_{1j}\;\frac{1}{1-z}, \nonumber \\
g_{2j}(z) & = & \delta_{2j} + z w_{21}\;g_{1j}(z) +
zw_{22}\;g_{2j}(z) + zw_{23}\;g_{3j}(z), \nonumber \\
g_{3j}(z) & = & \delta_{3j} + zw_{32}\;g_{2j}(z) +
zw_{33}\;g_{3j}(z) + zw_{34}\;g_{4j}(z), \nonumber \\
g_{4j}(z) & = & \delta_{4j} + w_{43}\;g_{3j}(z) +
zw_{44}\;g_{4j}(z) + zw_{45}\;g_{5j}(z), \nonumber \\
g_{5j}(z) & = & \delta_{5j}\;\frac{1}{1-z}. \nonumber
\end{eqnarray}
This can be simplified and rewritten in the form:
\begin{eqnarray}
(1 - zw_{22})\;g_{2j}(z) - zw_{23}\;g_{3j}(z) & = &  \delta_{2j} +
w_{21}\frac{z}{1-z}\;\delta_{1j},  \label{34} \\
-zw_{32}\;g_{2j}(z) + (1 - zw_{33})\;g_{3j}(z) -
zw_{34}\;g_{4j}(z) & = & \delta_{3j}, \label{35} \\
-zw_{43}\;g_{3j}(z) + (1 - w_{44})\;g_{4j}(z) & = & \delta_{4j} +
w_{45}\frac{z}{1-z}\;\delta_{5j} \label{36}.
\end{eqnarray}
After elementary algebra, we can determine all the generating
functions $g_{ij}(z), \;\; (i,j=1,2,3,4,5)$, nevertheless we are
now interested only in those functions which correspond to
processes starting from the state ${\mathcal S}_{3}$. In this case
we have
\begin{eqnarray}
g_{31}(z) & = & \frac{z^{2}}{1-z}\;\frac{w_{32} w_{21} (1 - w_{44}
z)}{D(z)}, \label{37} \\
g_{32}(z) & = & \frac{w_{32} z (1 - w_{44}(z) z)}{D(z)},
\label{38} \\
g_{33}(z) & = & \frac{(1 - w_{22} z)(1 - w_{44}
z)}{D(z)},
\label{39} \\
g_{34}(z) & = & \frac{w_{43} (1 - w_{22} z)}{D(z)}, \label{40} \\
g_{35}(z) & = & \frac{z^{2}}{1-z}\;\frac{w_{34} w_{45} (1 - w_{22}
z)}{D(z)}, \label{41}
\end{eqnarray}
where
\begin{equation} \label{42}
D(z) = (1 - w_{44} z)\left[(1 - w_{22} z)(1 - w_{33} z) - w_{23}
w_{32} z^{2}\right] - (1 - w_{22} z) w_{34} w_{43} z^{2}.
\end{equation}
Applying Tauber's Theorem we obtain that
\begin{equation} \label{43}
\lim_{n \rightarrow \infty} w_{31}(n) = \lim_{z \uparrow 1}
(1-z)\;g_{31}(z) = \frac{w_{32} w_{21} (1 - w_{44})}{D(1)},
\end{equation}
\begin{equation} \label{44}
\lim_{n \rightarrow \infty} w_{3j}(n) = \lim_{z \uparrow 1}
(1-z)\;g_{3j}(z) = 0, \;\;\;\;\;\; j = 2, 3, 4,
\end{equation}
and
\begin{equation} \label{45}
\lim_{n \rightarrow \infty} w_{35}(n) = \lim_{z \uparrow 1}
(1-z)\;g_{35}(z) = \frac{w_{34} w_{45} (1 - w_{22})}{D(1)}.
\end{equation}
Performing the substitutions
\[ w_{22} = 1 - w_{21} - w_{23}, \;\;\;\; w_{33} = 1 - w_{32} -
w_{34} \;\;\;\; w_{44} = 1 - w_{43} - w_{45}, \] we have
\begin{equation} \label{46}
w_{31}(\infty) = \frac{w_{32} w_{21}(w_{43} + w_{45})}{w_{34}
w_{45}(w_{21} + w_{23}) + w_{32} w_{21} (w_{43} + w_{45})},
\end{equation}
and
\begin{equation} \label{47}
w_{35}(\infty) = \frac{w_{34} w_{45}(w_{21} + w_{23})}{w_{34}
w_{45}(w_{21} + w_{23}) + w_{32} w_{21} (w_{43} + w_{45})}.
\end{equation}
It is elementary to show that
\begin{equation} \label{48}
F_{31} = \varphi_{31}(1) = w_{31}(\infty) \;\;\;\;\;\; \text{and}
\;\;\;\;\;\; F_{35} = \varphi_{35}(1) = w_{35}(\infty).
\end{equation}
In order to prove these equations, let us take into account
relationship (\ref{15}) and write
\[ \varphi_{31}(z) = \frac{\omega_{31}(z)}{1 + \omega_{11}(z)} =
\frac{g_{31}(z)}{g_{11}(z)}, \] and
\[ \varphi_{35}(z) = \frac{\omega_{35}(z)}{1 + \omega_{55}(z)} =
\frac{g_{35}(z)}{g_{55}(z)}. \] Since
\[ g_{11}(z) = g_{55}(z) = \frac{1}{1 -z}, \]
we have
\begin{equation} \label{49}
\varphi_{31}(z) = (1 - z) g_{31}(z) \;\;\;\;\;\; \text{and}
\;\;\;\;\;\; \varphi_{35}(z) = (1 - z) g_{35}(z).
\end{equation}
Comparing (\ref{43}) and (\ref{45}) with (\ref{49}) we see that
Eqs. (\ref{48}) are true.

It is convenient to write the absorption probabilities $F_{31}$
and $F_{35}$ in the form:
\begin{equation} \label{50}
F_{31} = \frac{1}{1 + r}, \;\;\;\;\;\; \text{and} \;\;\;\;\;\;
F_{35} = \frac{r}{1 + r},
\end{equation}
where
\begin{equation} \label{51}
r = \frac{w_{34} w_{45} (w_{21} + w_{23})}{w_{32} w_{21} (w_{43} +
w_{45})}
\end{equation}
and we see immediately that $F_{31} + F_{35} = 1$, as expected.

It seems to be worthwhile to study the history of a particle
starting its random walk from the state ${\mathcal S}_{3}$.
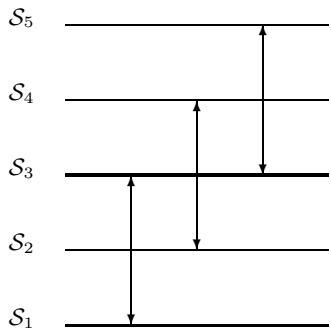
\begin{figure}[ht!]  \setlength\unitlength{0.5cm}
\begin{center}
\begin{picture}(8,12)  \put(0.5,
1){\makebox{${\mathcal S}_{1}$}}\put(2,1){\line(1,0){7}} \put(0.5,
3){\makebox{${\mathcal S}_{2}$}}\put(2,3){\line(1,0){7}}
 \put(0.5, 5){\makebox{${\mathcal
S}_{3}$}}\put(2,5){\thicklines  \line(1,0){7}} \put(0.5,
7){\makebox{${\mathcal  S}_{4}$}}\put(2,7){\line(1,0){7}}
\put(0.5, 9){\makebox{${\mathcal S}_{5}$}}\put(2,9){\line(1,0){7}}
\put(5.5, 5){\vector(0,1){2}}  \put(5.5, 5){\vector(0,-1){2}}
\put(3.75, 3){\vector(0,1){2}}  \put(3.75, 3){\vector(0,-1){2}}
\put(7.25, 7){\vector(0,1){2}}  \put(7.25, 7){\vector(0,-1){2}}
\end{picture}
\end{center}  \caption{\label{f1} {\footnotesize Illustration of
the random walk on a ladder of five rungs}}
\end{figure}
Let us consider a trap containing a special ladder with $5$ rungs.
Each rung corresponds to a given state of the Markov chain under
investigation. The process starts when a particle enters (say,) on
the third rung of the ladder, i.e., in the state ${\mathcal
S}_{3}$. Once the particle has entered, it is free to move up and
down the rungs randomly . Fig. 1 illustrates this random walk. If
the particle reaches the states either ${\mathcal S}_{1}$ or
${\mathcal S}_{5}$, it is absorbed. (If the random walk is
considered as a game, then the absorption state with probability
smaller than $1/2$ is the ``winning" state.) Having chosen the
transition matrix
\begin{equation} \label{52}
\mathbf{w} =
\begin{pmatrix}
1 & 0 & 0 & 0 & 0 \\ 4/36 & 24/36 & 8/36 & 0 & 0 \\
0 & 5/36 & 29/36 & 2/36 & 0 \\
0 & 0 & 4/36 & 24/36 & 8/36 \\
0 & 0 & 0 & 0 & 1
\end{pmatrix},
\end{equation}
\begin{figure} [ht!] \protect \centering{
\includegraphics[height=8cm, width=10cm]{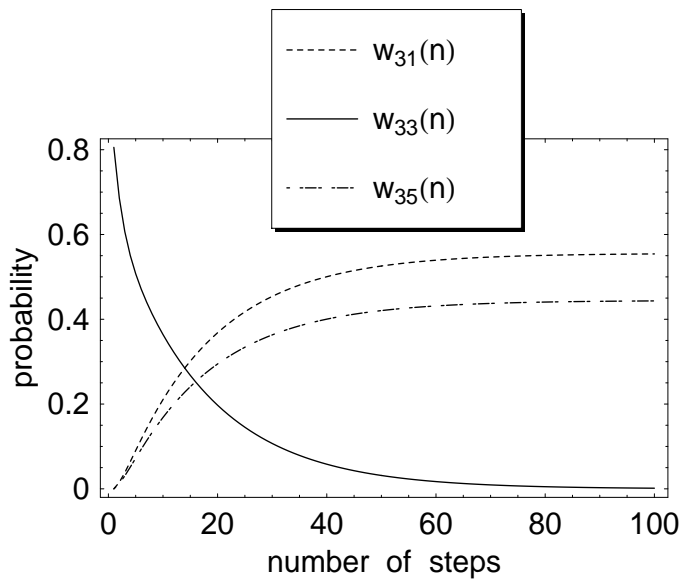}}\protect
\vskip -0.3cm \protect \caption{\label{f2}{\footnotesize
Dependencies of the probabilities $w_{31}(n), w_{33}(n)$ and
$w_{35}(n)$ on the number of steps $n$}}
\end{figure}
we calculated the dependencies of  probabilities $w_{31}(n),
w_{33}(n)$ and $w_{35}(n)$ on the number of steps $n$. The results
of calculation are shown in Fig. 2. We see that the probability to
find the particle after $n \approx 100$ steps in the transient
state ${\mathcal S}_{3}$ is practically zero. The same holds for
the transient states ${\mathcal S}_{2}$ and ${\mathcal S}_{4}$.
After $n \approx 100$ steps the particle is absorbed either in
${\mathcal S}_{1}$ with probability $w_{31}(100) \approx F_{31} =
5/9$ or in ${\mathcal S}_{5}$ with probability $w_{35}(100)
\approx F_{35} = 4/9$.

It is instructive to determine also the probabilities $F_{32},
F_{33}$ and $F_{34}$. As a reminder, we note that $F_{3j}$ is the
probability that a particle starting from ${\mathcal S}_{3}$
passes through ${\mathcal S}_{j}, \;\; (j=2,3,4)$ \textit{at least
once.} By using the transition matrix (\ref{52}) we obtain the
following values: $F_{32} = 15/19, \; F_{33} = 11/12$ and $F_{34}
= 6/11$. Fig. 3 shows the histogram of these probabilities.
\begin{figure} [ht!] \protect \centering{
\includegraphics[height=8cm, width=12cm]{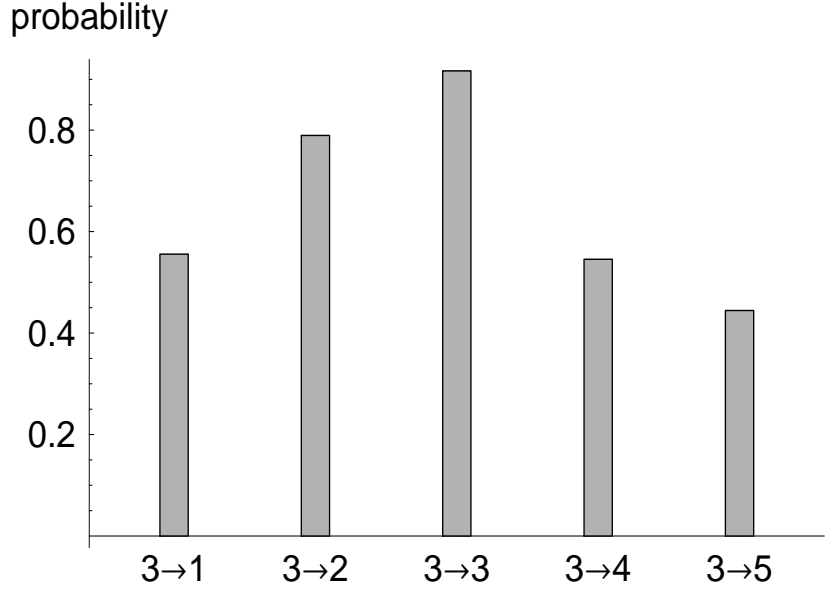}}\protect
\vskip -0.3cm \protect \caption{\label{f3}{\footnotesize
Probabilities that a particle passes through the state ${\mathcal
S}_{j}, \;\; (j=1,2,3,4,5)$ at least once provided that it started
from ${\mathcal S}_{3}$ }}
\end{figure}
It is evident that passing through either ${\mathcal S}_{1}$ or
${\mathcal S}_{5}$ at least once means that the particle is
absorbed. As expected in the present case, the probability
$F_{33}$ that the particle starting from ${\mathcal S}_{3}$
returns to ${\mathcal S}_{3}$ at least once, is nearly $1$. It is
to mention that the two absorbing states ${\mathcal S}_{1}$ and
${\mathcal S}_{5}$ are recurrent since $F_{11} = F_{55} = 1$.

In what follows we would like to deal with the
\textit{determination of the absorption time probability}. Denote
by $\tau_{i}$ the number of steps leading to the absorption of a
particle starting its random walk from the state ${\mathcal
S}_{i}$. By definition, $f_{i1}(n)$ and $f_{i5}(n)$ are the
probabilities that the particle starting from the state ${\mathcal
S}_{i}, \;\; (i=2,3,4)$ is absorbed exactly at the $n$th step in
${\mathcal S}_{1}$ or in ${\mathcal S}_{5}$, respectively. Hence
we can write that
\begin{equation} \label{53}
{\mathcal P}\{\tau_{i}=n\} = T_{i}(n) = f_{i1}(n) + f_{i5}(n),
\;\;\;\;\;\; i=2,3,4.
\end{equation}
It is easy to prove that
\begin{equation} \label{54}
T_{i}(n) = w_{i1}(n) - w_{i1}(n-1) + w_{i5}(n) - w_{i5}(n-1),
\;\;\;\;\;\; \forall \; n \geq 1.
\end{equation}
From (\ref{12}) one obtains
\begin{eqnarray}
w_{i1}(n) & = & \sum_{k=1}^{n} f_{i1}(k)\;w_{11}(n-k), \nonumber \\
w_{i5}(n) & = & \sum_{k=1}^{n} f_{i5}(k)\;w_{55}(n-k), \nonumber
\end{eqnarray}
and by taking into account that
\[ w_{11}(\ell) = w_{55}(\ell) = 1, \;\;\;\;\;\; \forall \; \ell
\geq 0, \] one has
\[ w_{i1}(n) = \sum_{k=1}^{n} f_{i1}(k) \;\;\;\;\;\;
\text{and} \;\;\;\;\;\; w_{i5}(n) = \sum_{k=1}^{n} f_{i5}(k).
\] It follows immediately from these equations that
\[ f_{i1}(n) = w_{i1}(n) - w_{i1}(n-1) \;\;\;\;\;\; \text{and}
\;\;\;\;\;\; f_{i5}(n) = w_{i5}(n) - w_{i5}(n-1), \] and this
completes the proof. The absorption time probabilities $T_{i}(n),
\;\; (i=2,3,4)$ can be determined by the ``forward" equations:
\[ f_{i1}(n) = \sum_{\ell=2}^{4} w_{i \ell}(n-1)\;w_{\ell 1} \]
and
\[ f_{i5}(n) = \sum_{\ell=2}^{4} w_{i \ell}(n-1)\;w_{\ell 5}. \]
By using these expressions one can write
\begin{equation} \label{55}
T_{i}(n) = \sum_{\ell=2}^{4} w_{i \ell}(n-1)\left[\;w_{\ell 1} +
\;w_{\ell 5}\right],
\end{equation}
which in the case of $\mathbf{w}$ defined by (\ref{27}) has the
following form:
\begin{equation} \label{56}
T_{i}(n) = w_{i2}(n-1)\;w_{21} + w_{i4}(n-1)\;w_{45}.
\end{equation}
For the sake of completeness, we would like to show that
\begin{equation} \label{57}
\sum_{n=1}^{\infty} T_{i}(n) = 1.
\end{equation}
In the case of Eq. (\ref{53}) we see that
\[ \sum_{n=1}^{\infty} T_{i}(n) = F_{i1} + F_{i5}, \]
and by using the expression (\ref{26}) we find (\ref{57}). In the
case of Eq. (\ref{55})
\[ \sum_{n=1}^{\infty} T_{i}(n) = \sum_{\ell=2}^{4}
\left[\sum_{n=1}^{\infty}w_{i \ell}(n-1) \right] \;
\left[\;w_{\ell 1} + \;w_{\ell 5}\right] = \]
\[ \sum_{\ell=2}^{4} \left[\delta_{i \ell} +
\omega_{i \ell}(1)\right]\;\left[\;w_{\ell 1} + \;w_{\ell
5}\right] = \sum_{\ell=2}^{4} g_{i \ell}(1)\;\left[\;w_{\ell 1} +
\;w_{\ell 5}\right] = F_{i1} + F_{i5} = 1. \]

\begin{figure} [ht!] \protect \centering{
\includegraphics[height=8cm, width=12cm]{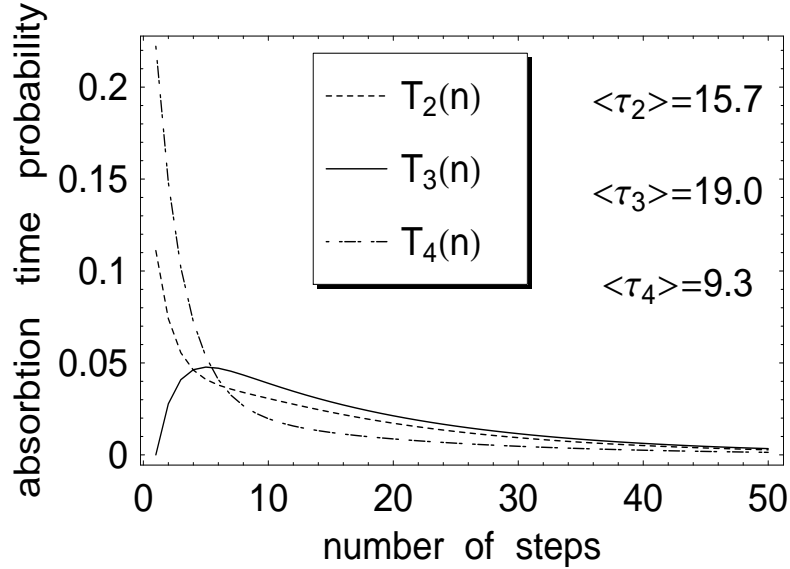}}\protect
\vskip -0.3cm \protect \caption{\label{f4}{\footnotesize
Absorption time probabilities of a particle starting its random
walk from the state ${\mathcal S}_{i}, \;\; (i=2,3,4)$ }}
\end{figure}

Using the transition matrix $\mathbf{w}$ given by (\ref{52}), we
calculated the dependence of the probability $T_{i}(n)$ on the
number of steps $n$. The results are seen in Fig. 4. As expected,
if the starting state is ${\mathcal S}_{3}$, then the probability
$T_{3}(n)$  varies differently with the step number as the
probabilities $T_{2}(n)$ and $T_{4}(n)$. It is characteristic the
probabilities have a rather long tail. Since $T_{i}(n)$ is the
probability that a particle starting from ${\mathcal S}_{i}$ is
absorbed exactly in the $n$th step, the expectation and the
standard deviation of the absorption time $\tau_{i}$ are given by
\begin{equation} \label{58}
\mathbf{E}\{\tau_{i}\} = \sum_{n=1}^{\infty} n\;T_{i}(n) =
<\tau_{i}>,
\end{equation}
and
\begin{equation} \label{59}
\mathbf{D}\{\tau_{i}\} = \left[\sum_{n=1}^{\infty} (n -
<\tau_{i}>)^{2}\;T_{i}(n)\right]^{1/2}.
\end{equation}
For a transition matrix of the form (\ref{52}) these values are
presented in the Table I.

\begin{table}[ht!]
\caption{\label{t1} Expectations and the standard deviations of
the absorption time}
\begin{ruledtabular}
\begin{tabular}{|c|c|c|c|}
& ${\mathcal S}_{2}$ & ${\mathcal S}_{3}$ & ${\mathcal S}_{4}$ \\
\hline $\mathbf{E}\{\tau_{i}\}$ & 15.7 & 19.0 & 9.3 \\ \hline
$\mathbf{D}\{\tau_{i}\}$ & 16.3 & 16.4 & 13.3 \\
\end{tabular}
\end{ruledtabular}
\end{table}

\subsection{Properties of the absorption probability $\mathbf{F_{31}}$}

As it has been shown, $F_{31}$ is the probability that a particle
starting its random walk from the state ${\mathcal S}_{3}$ is
finally absorbed in the state ${\mathcal S}_{1}$.~\footnote{There
is no need to deal separately with the absorption probability
$F_{35}$ since $F_{31} + F_{35} = 1$.} If $F_{31}
> 1/2$, then ${\mathcal S}_{1}$ is called a ``losing" state, while if
$F_{31} < 1/2$, then it is a ``winning" state. The game is ``fair"
when $F_{31} = 1/2$, i.e. when the equation
\begin{equation} \label{60}
w_{32} w_{21} \left(w_{43} + w_{45}\right)  = w_{34} w_{45}
\left(w_{21} + w_{23}\right)
\end{equation}
is fulfilled as it follows from Eq. (\ref{51}).

Astumian \cite{ast01} proposed two transition matrices, namely

\[
\mathbf{w_{1}} =
\begin{pmatrix}
1 & 0 & 0 & 0 & 0 \\ 4/36 & 24/36 & 8/36 & 0 & 0 \\
0 & 5/36 & 29/36 & 2/36 & 0 \\
0 & 0 & 4/36 & 24/36 & 8/36 \\
0 & 0 & 0 & 0 & 1
\end{pmatrix} \;\;\;\;\;\; \mathbf{w_{2}} =
\begin{pmatrix}
1 & 0 & 0 & 0 & 0 \\ 4/36 & 24/36 & 8/36 & 0 & 0 \\
0 & 5/36 & 29/36 & 2/36 & 0 \\
0 & 0 & 4/36 & 24/36 & 8/36 \\
0 & 0 & 0 & 0 & 1
\end{pmatrix}  \]
resulting in the absorption probability $F_{31} = 5/9 > 1/2$ and
showed that the arithmetic mean of these two matrices
\[ \mathbf{w} = \frac{1}{2}\left(\mathbf{w_{1}} +  \mathbf{w_{1}}\right) = \]
\[ = \begin{pmatrix}
1 & 0 & 0 & 0 & 0 \\ 9/72 & 53/72 & 10/72 & 0 & 0 \\
0 & 9/72 & 53/72 & 10/72 & 0 \\
0 & 0 & 9/72 & 53/36 & 10/72 \\
0 & 0 & 0 & 0 & 1
\end{pmatrix} \]
brings about the probability $F_{31} = 9/19 < 1/2$, i.e., in this
case the state ${\mathcal S}_{1}$ becomes ``winning" state. This
property of the transition matrix (\ref{27}) is general \textit{if
the diagonal entries of the matrix are different from zero}. By
using a simple example we would like to demonstrate this
statement.

Let us choose the transition matrix in the following form:
\begin{equation} \label{61}
\mathbf{w} =
\begin{pmatrix}
1 & 0 & 0 & 0 & 0 \\ a & 1-a-b & b & 0 & 0 \\
0 & b & 1-a-b-x & a+x & 0 \\
0 & 0 & a+x & 1-a-b & b-x \\
0 & 0 & 0 & 0 & 1
\end{pmatrix}.
\end{equation}
One obtains immediately that
\begin{equation} \label{62}
F_{31} = H(x) = \frac{a b}{a b + (a+x)(b-x)},
\end{equation}
where \[ 0 < a < 1, \;\;\;\; 0 < b < 1, \;\;\;\; 0 < a + b < 1
\;\;\;\;\;\; \text{and} \;\;\;\;\;\; -a < x < \min(b, 1-a-b). \]
If $x = 0$ or $x = b - a$, then  the game is ``fair", i.e.,
$F_{31} = 1/2.$ The function $H(x)$ assumes its minimal value at
\[ x = x_{min} = \frac{1}{2}(b-a), \] and this value is
\[ H(x_{min}) = \left\{ \begin{array} {ll}
1/2, & \text{if $a=b$,} \\
\text{ } & \text{ } \\
\frac{4 a b}{4 a b + (a + b)^{2}} < 1/2, & \text{if $a \neq b$.}
\end{array} \right. \] Introducing the notation $x = x_{min} + y$
one has
\[ H(x_{min}+y) = J(y) = \frac{4 a b}{4 a b  + (a + b)^{2} - 4
y^{2}}. \] Choosing  $y$ according to the inequalities \[ x_{1} =
x_{min} + y
> 0 \;\;\;\;\;\; \text{and} \;\;\;\;\;\;x_{2} = x_{min} - y < b -
a, \] i.e., $y > \vert x_{min} \vert$ and $a \neq b$ one finds
that \[ H(x_{1}) = H(x_{2}) > \frac{1}{2} \;\;\;\;\;\; \text{and}
\;\;\;\;\;\; H\left(\frac{x_{1}+x_{2}}{2}\right) < \frac{1}{2}. \]
Evidently, \textit{there are infinitely many pairs of transition
matrices which result in probabilities of losing in the state
${\mathcal S}_{1}$ but the arithmetic means of corresponding pairs
bring about probabilities of winning in the state ${\mathcal
S}_{1}$}.

\begin{figure} [ht!] \protect \centering{
\includegraphics[height=8cm, width=12cm]{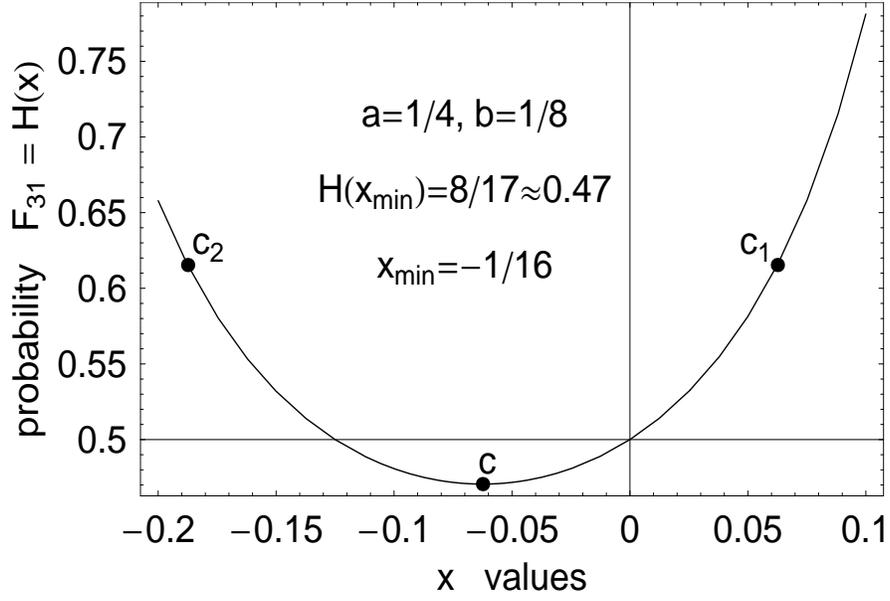}}\protect
\vskip -0.3cm \protect \caption{\label{f5}{\footnotesize
Dependence of $F_{31}= H(x)$ on the parameter $x$ of the
transition matrix (\ref{61}) when $a=1/4$ and $b=1/8$ }}
\end{figure}

For the sake of illustration in Fig. 5 the probability $F_{31} =
H(x)$ vs. $x$ curve is plotted by the values $a = 1/4$ and $b =
1/8$. The black points $c_{1},\; c_{2}$ and $c$ correspond to the
probabilities \[ H(x_{1}=1/16) = H(x_{2}=-3/16) = 8/13
\;\;\;\;\;\; \text{and} \;\;\;\;\;\;
H\left[\frac{1}{2}(x_{1}+x_{2})=-1/16\right] = 8/17,
\] respectively. It seems to be not superfluous to write down the
corresponding transition matrices:
\begin{equation} \label{63}
\mathbf{w_{1}} =
\begin{pmatrix}
1 & 0 & 0 & 0 & 0 \\ 4/16 & 10/16 & 2/16 & 0 & 0 \\
0 & 2/16 & 9/16 & 5/16 & 0 \\
0 & 0 & 5/16 & 10/16 & 1/16 \\
0 & 0 & 0 & 0 & 1
\end{pmatrix}, \;\;\;\;\;\; \mathbf{w_{2}} =
\begin{pmatrix}
1 & 0 & 0 & 0 & 0 \\ 4/16 & 10/16 & 2/16 & 0 & 0 \\
0 & 2/16 & 13/16 & 1/16 & 0 \\
0 & 0 & 1/16 & 10/16 & 5/16 \\
0 & 0 & 0 & 0 & 1
\end{pmatrix},
\end{equation}
and
\begin{equation} \label{64}
\mathbf{w} = \frac{1}{2}(\mathbf{w_{1}} + \mathbf{w_{2}}) =
\begin{pmatrix}
1 & 0 & 0 & 0 & 0 \\ 4/16 & 10/16 & 2/16 & 0 & 0 \\
0 & 2/16 & 11/16 & 3/16 & 0 \\
0 & 0 & 3/16 & 10/16 & 3/16 \\
0 & 0 & 0 & 0 & 1
\end{pmatrix}.
\end{equation}

By choosing $y$ values in the allowed interval, we can construct
infinitely many transition matrices with just described
properties.

\begin{figure} [ht!] \protect \centering{
\includegraphics[height=8cm, width=12cm]{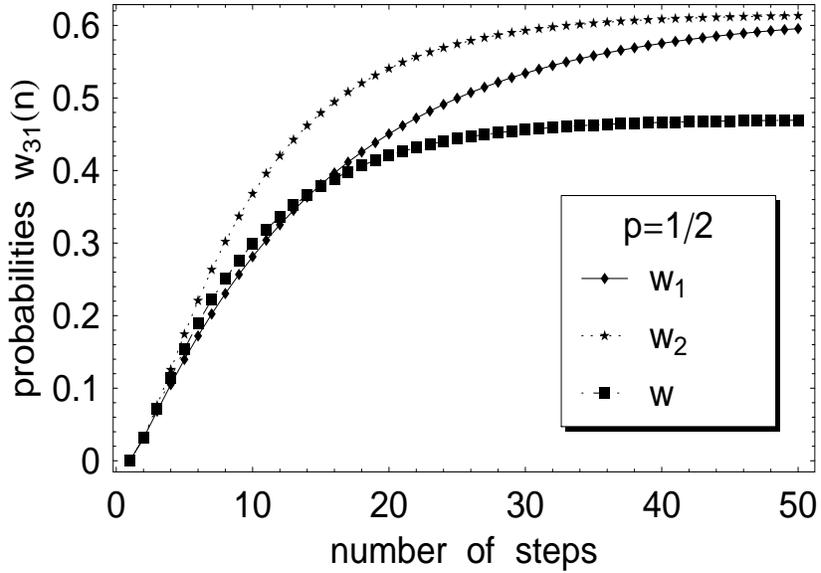}}\protect
\vskip -0.3cm \protect \caption{\label{f6}{\footnotesize
Dependence of $w_{31}(n)$ on the number of steps $n$ in the case
of $p=1/2$ }}
\end{figure}

\begin{figure} [ht!] \protect \centering{
\includegraphics[height=8cm, width=12cm]{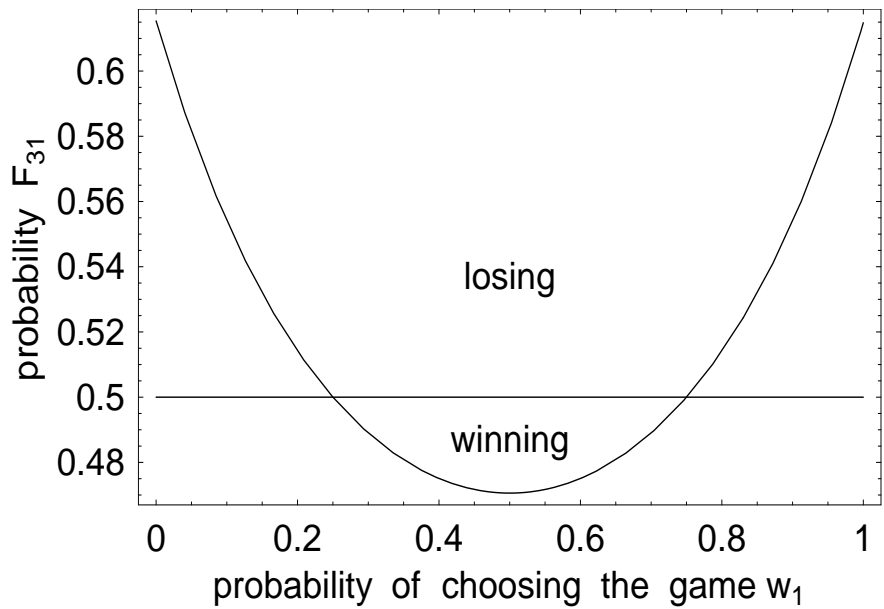}}\protect
\vskip -0.3cm \protect \caption{\label{f7}{\footnotesize
Dependence of the absorption probabilities $F_{31}(n)$ on the
chance $p$ moving the particle at a given step according to the
transition matrix $\mathbf{w_{1}}$ }}
\end{figure}

Let us now define a Markov chain with transition matrix
$\mathbf{w}$ randomly chosen from $\mathbf{w_{1}}$ and
$\mathbf{w_{2}}$ defined by (\ref{63}). In this case
\[ \mathbf{w}(n) = \left[p \mathbf{w_{1}} + (1-p)
\mathbf{w_{2}}\right] \times \mathbf{w}(n-1), \] i.e.,
\begin{equation} \label{65}
\mathbf{w}(n) = \left[p \mathbf{w_{1}} + (1-p)
\mathbf{w_{2}}\right]^{n}.
\end{equation}

In Fig. 6 the dependencies of the absorption probabilities
$w_{31}(n)$~\footnote{As seen before, $w_{31}(n)$ is the first
entry of the third row of the matrix $\mathbf{w}(n)$.} on the
number of steps $n$ are shown when the transition matrices are
$\mathbf{w_{1}}, \; \mathbf{w_{2}}$ and $\mathbf{w}$,
respectively. The last one corresponds to the random selection of
the entries from $\mathbf{w_{1}}$ and $\mathbf{w_{2}}$ with
probability $p=1/2$. Obviously, not all values of $p \in [0,1]$
bring about a  ``winning" game, i.e., an absorption probability
less than $1/2$.

Taking into account the transition matrices $\mathbf{w_{1}}$ and
$\mathbf{w_{2}}$ defined by (\ref{63}), we determined the
dependence of $F_{31}$ on $p$. As seen in Fig. 7, there is a well
defined subinterval $[p_{1},p_{2}] \in [0,1]$ containing the $p$
values which result in absorption probabilities $F_{31}$ smaller
than $1/2$. In the present case we obtained that $p_{1} = 0.25$
and $p_{2} = 0.75$.

\section{Conclusions}

It has been shown that the random walk of a particle defined by
the stochastic transition  matrix of a Markov chain is equivalent
to an Astumian type game if the diagonal entries of the matrix are
different from zero and the first $(w_{11})$ as well as the last
$(w_{NN})$ entries are equal to $1$. By using a simple example, we
have proved that there are infinitely many pairs of transition
matrices which result in absorption probabilities in the state
${\mathcal S}_{1}$ larger than $1/2$ but the arithmetic means of
the corresponding pairs lead to probabilities smaller than $1/2$.

\end{document}